\documentclass[prl,aps,twocolumn, floatfix, superscriptaddress,showpacs]{revtex4}
\usepackage{amsmath,bm,graphicx}
\usepackage{amssymb}
\usepackage{amsfonts}
\usepackage{times}

\usepackage{graphicx,color}
\definecolor{r}{rgb}{1,0,0}
\definecolor{b}{rgb}{0,0,1}

\begin{document}

\title{Convective instability and mass transport of diffusion layers in a Hele-Shaw geometry}

\author{Scott Backhaus}
\affiliation{MPA-CMMS, Los Alamos National Laboratory, Los Alamos, NM 87545, USA}
\email[Corresponding Author: ]{backhaus@lanl.gov}

\author{Konstantin Turitsyn}
\affiliation{Theoretical Division and Center for Nonlinear Studies, Los Alamos National Laboratory, Los Alamos, NM 87545, USA}

\author{R. E. Ecke}
\affiliation{Center for Nonlinear Studies, Los Alamos National Laboratory, Los Alamos, NM 87545, USA}

\date{\today}

\begin{abstract}
We consider experimentally the instability and mass transport of a porous-medium flow in a Hele-Shaw geometry. In an initially stable configuration, a lighter fluid (water) is located over a heavier fluid (propylene glycol). The fluids mix via diffusion with some regions of the resulting mixture being heavier than either pure fluid. Density-driven convection occurs with downward penetrating dense fingers that transport mass much more effectively than diffusion alone. We investigate the initial instability and the quasi steady state. The convective time and velocity scales, finger width, wave number selection, and normalized mass transport are determined for $6,000 < Ra < 90,000$. The results have important implications for determining the time scales and rates of dissolution trapping of carbon dioxide in brine aquifers proposed as possible geologic repositories for sequestering carbon dioxide.
\end{abstract}

\maketitle

Convective processes that transport heat and mass are critical to many natural phenomena in the atmosphere and oceans as well as in planetary and stellar interiors \cite{Ahlers2009review}. Another important example of current scientific and technological interest is the convective mixing of supercritical carbon dioxide sequestered in a brine-saturated porous medium, \cite{IPCC2005} i.e., a deep saline aquifer. When injected into such aquifers, the low density CO$_2$ rises to the top of the aquifer where it is contained by cap rock formations. As a pure phase, the buoyancy of the supercritical CO$_2$ may drive it back to the surface defeating attempts to sequester it. Diffusion into the underlying brine results in a denser mixture eliminating buoyancy, but this process is very slow and the pure phase of supercritical CO$_2$ would remain mostly intact for long times ($\approx$ 10,000 years) unless other processes occur.

One scenario for enhanced dissolution of the CO$_2$ relies on this unusual property of the CO$_2$-brine system, i.e. CO$_2$-brine mixtures are denser than pure brine\cite{IPCC2005} making gravitationally-driven convection is possible.  However, the diffusion layer must thicken enough to become unstable, a process similar to the thickening and subsequent instability of thermal layers \cite{Foster65}. Analytical calculations and numerical simulations have been performed \cite{EnnisKing2005,Chen2006,Riaz2006,EnnisKing2007,Rapaka2008,Rapka2009} to determine the instability incubation time and the subsequent mass transport in CO$_2$-brine systems. Corresponding laboratory experiments\cite{Pruess2010}, however, have not succeeded in capturing the essential features of such diffusion-driven, convectively unstable layers.

The difficulties in realizing a laboratory experiment for the purposes of testing the scenario of unstable diffusion layers in porous media are considerable, e.g. establishing the stable initial conditions assumed by calculations and simulations. Also, direct experiments using supercritical CO$_2$ and brine are complicated by high pressure ($\approx$100 bar). Instead, we have developed an atmospheric-pressure analog fluid system that mimics many features of the CO$_2$-brine system. It consists of water located vertically above propylene glycol (PPG) in a Hele-Shaw cell. As an analog system, water replaces the CO$_2$ and propylene glycol replaces the brine. In this cell, we can create relatively unperturbed initial conditions such that the instability occurs uniformly over a wide lateral portion of the cell. We measure the temporal evolution of the system using optical shadowgraph and determine the wave number selection, the time and velocity scales, the finger width, and the mass transport rate. Our results establish the basic description of diffusively-driven convective instability in such systems.

\begin{figure}
\begin{center}
\includegraphics[width = 2.5 in]{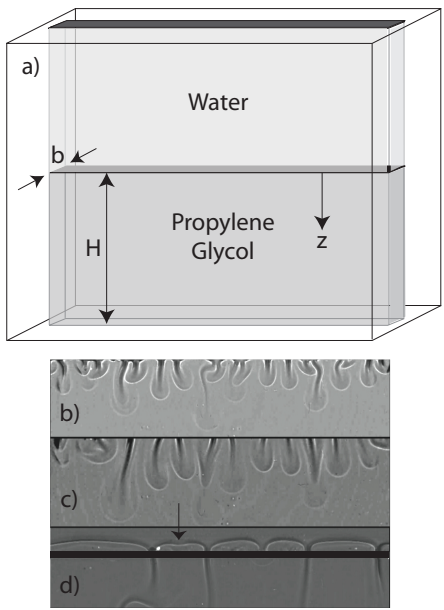}
\vspace{-0.5cm}
\end{center}
\caption{a) Schematic illustration of the Hele-Shaw cell where gap width $b$, PPG reservoir height $H$, and vertical coordinate $z$ are indicated. A thin metal shim separating the two fluids is withdrawn horizontally to begin an experimental run. Images at three intermediate times $t^*$ are shown for a run with K=$1.9\times 10^{-4}\ {\rm cm}^2$ and Ra=87,000:  b) $t^* = 2.0$, c) $t^* = 3.0$ , d) $t^* = 10.0$. Only a portion of the vertical extent of the image is shown. As seen in d), light scattering off of the metallic shims makes the precise location of the interface uncertain.  The vertical arrow in d) indicates the nucleation site of a new finger.
 }
 \label{fig:cell}
\end{figure}

The experimental apparatus in Fig.~\ref{fig:cell}, consists of a vertically-oriented Hele-Shaw cell with two 1.2-cm-thick polycarbonate sheets separated by steel shims of thickness $b$. The flow in the narrow gap $b$ between the parallel surfaces obeys Darcy's Law with an isotropic permeability $K = b^2/12$ and a porosity of unity. We vary $b$ from 0.25 to 0.48 mm giving values of $K$ in the range 5,000-20,000 Darcy (i.e. $0.54 \times 10^{-4} \le K \le 1.94 \times 10^{-4} {\rm cm}^2$). The lateral dimension of the flow region is 7.6 cm, and the initial depth of the bottom fluid $H$ ranges from 1.25 to 5 cm. Water ($\rho_w = 1.000$ g/cc) is lighter than PPG ($\rho_{ppg} = 1.035$ g/cc\cite{dowppgdata}) and is initially above the PPG. Water-PPG mixtures with a water concentration $c_w$ between 0 and $\approx$ 50\% have $\rho \geq \rho_{PPG}$. After a period of diffusive growth, the mixed fluid interface may become gravitationally unstable, similar to the proposed instability when supercritical CO$_2$ dissolves into brine. There are, however, some important differences between the two fluid systems. Water and PPG are fully miscible whereas CO$_2$ is partially miscible in brine with a low saturation concentration so that the transport properties of the CO$_2$-brine system do not depend strongly on the CO$_2$ concentration. In contrast, the viscosity and mass diffusivity of the water-PPG system are highly dependent on concentration. When computing experimental parameters or making comparisons to calculations or simulations, we use the transport properties of a 30\% water-70\% PPG mixture, i.e., $\nu = 0.1\ {\rm cm}^2/{\rm s}$ \cite{dowppgdata} and $D = 1 \times 10^{-6}\ {\rm cm}^2/{\rm s}$ (estimated using data from \cite{D_ppg}), which corresponds to the mixture of maximum density $\rho$=1.044 g/cc or the maximum density difference of $\Delta \rho$ = 0.009 g/cc relative to the underlying PPG.

The upper and lower sections of the Hele-Shaw cell are separated by a system of thin shims of maximum thickness 0.01 cm. The upper and lower reservoirs are filled with pure water and pure PPG, respectively. A 0.0076-cm-thick ``shutter shim'' is slowly removed horizontally from between the two fluids leaving behind the 0.01-cm-thick ``frame shim'' to ensure a liquid-tight seal. Small perturbations of the fluid-fluid interface were observed during some runs, but these did not seem to influence the flow appreciably for measurements presented here. The removal of the shutter shim opens up a narrow 0.01-cm-thick horizontal gap at the interface between the upper and lower cells. Our cell filling method ensures that this pocket is occupied by water after the shim removal \cite{Note-drain}.

The development of the flow is visualized using optical shadowgraph which images the mixture concentration via its index of refraction variation. The images resolve an area of lateral size $L = 3.0$ cm and height 2.4 cm with 1280 $\times$ 1024 pixels in the horizontal and vertical directions, respectively. Approximately 2000 frames are gathered during each run with the camera frame rate dependent on the permeability of the cell. (Total run times range between 2.5 and 10 hours). The flow develops through several stages. Initially, the PPG preferentially diffuses into the water owing to an asymmetric dependence of $D(C)$ \cite{D_ppg}. As the dense region in the diffusion layer thickens and the influence of buoyancy overcomes lateral diffusion, hydrodynamic instability occurs as small perturbations in the dense diffusive interface grow into downward moving fingers. Figure \ref{fig:cell}b-d show the evolution of a typical experimental run at various times after initiation.

\begin{figure}
\includegraphics[width=2.5 in]{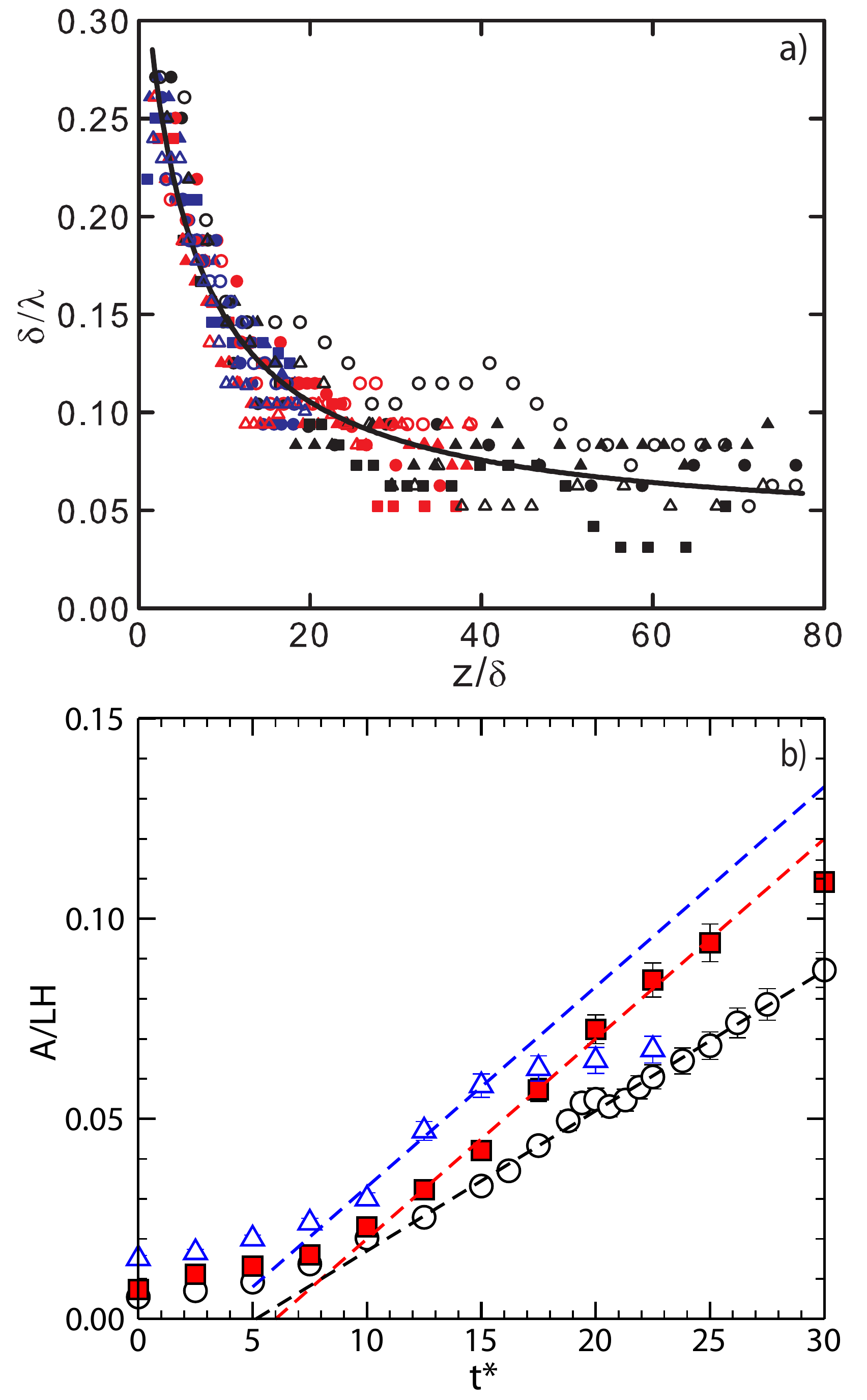}
\vspace{-0.3cm}
\caption{(Color online) a) $\delta/\lambda$ vs $z/\delta$. Different values of $H$ are $H=1.27$ cm (blue), $H=2.54$ cm (red), and $H=5.08$ cm (black).  The different symbols within each color indicate different values of $K$.  Over the range of the experiments in this work, $\delta$ is found to be roughly the same for all runs with an average value of $\delta=0.065$ cm.  The solid line is the best fit of the form $\delta/\lambda=(1.90+0.038 z/\delta)/(5.30+ z/\delta)$.  b) Area of water consumed $A/(LH)$ vs $t^*$ for several data sets with
$K = 0.77 \times 10^{-4}\ {\rm cm}^2$ and $Ra = 8,700$ (open triangles), $Ra = 17,400$ (solid squares), and $Ra = 34,800$ (open circles). The small offset in the initial $A/LH$ is due to the uncertainty in the location of the initial fluid-fluid interface as described in caption  of Fig.\ \ref{fig:cell}. }
\label{k-A-time}
\end{figure}

The equations governing the conservation of momentum, mass, and concentration; their various non-dimensional scalings; and solution techniques have been described in detail elsewhere \cite{EnnisKing2005,Chen2006,Riaz2006,EnnisKing2007,Rapaka2008,Rapka2009}. Here, we discuss the dimensionless parameters and point out a few details relevant to our experiment. Under the conditions of the experiment, the fluids and their mixture are incompressible. Conservation of momentum is adequately described by a two-dimensional Darcy's law because the Reynolds numbers $Re_b$ based on $b$ and the maximum observed velocities are less than 10$^{-3}$. The evolution of the mixture concentration is governed by the advection-diffusion equation, although the molecular diffusion coefficient $D$ may be modified by Taylor dispersion \cite{Taylor1953} in certain regions of higher flow velocity. For example, near the tips and within the core of fingers which move downward with velocity $v_f$, the Peclet number $Pe=b v_f/D\approx 50$ increasing the effective lateral diffusion coefficient by a factor 15. Elsewhere, the fluid velocities are significantly lower, and the effect on $D$ is negligible.

Recalling that $Re_b\ll 1$ and neglecting the impact of $Pe$ within the finger core, the only dimensionless number that governs the flow is the Rayleigh number which we define in the same way as \cite{EnnisKing2005,Chen2006,Riaz2006,EnnisKing2007,Rapaka2008,Rapka2009}, i.e., $Ra=\Delta \rho g K H/\rho\nu D$ where $g$ is the acceleration due to gravity and, as described above, all fluid properties are evaluated at maximum mixture density. Next, we define characteristic length, velocity, and time scales. The vertical convective motion of a blob of maximum density fluid is given by a balance between buoyancy and viscous drag, i.e., $v_c=\Delta \rho g K/\rho \nu$. At $v_c$, the time required for this blob to fall the entire depth $H$ of the lower fluid in Fig.\ \ref{fig:cell} is $t_c=H/v_c=H /(\Delta \rho gK/\rho \nu)$. We define $t^*=t/t_c$. Comparative time and velocity scales are the diffusion time $t_d=H^2/D$, i.e., the approximate time  to establish a diffusion gradient over $H$, and the diffusion velocity $v_d=D/H$. From these scaling relations, we see that $Ra = t_d/t_c=v_c/v_d$ describes the importance of convection relative to diffusion over the vertical distance $H$.

First, we consider the finger dynamics from instability onset to steady state (i.e. Fig.~\ref{fig:cell}b to d). After onset, several of the initially shorter fingers stagnate and disappear which we interpret as the longer dominant fingers out competing the shorter fingers and preferentially draining dense mixture from the diffusion interface. Later, the remaining dominant fingers move laterally along the interface, collide, and merge resulting in the coarse pattern of Fig.~\ref{fig:cell}d. When plotted versus the scaled depth $z/\delta$ of the longest finger, $\delta/\lambda$ shows consistent behavior among all the experimental runs (see Fig.~\ref{k-A-time}a). Here, $\delta$ is the measured finger width. Near $z/\delta\approx 1$, i.e. near the instability onset, $\delta/\lambda$ ranges between 0.22 and 0.27 consistent with the onset value of $\delta/\lambda\approx 0.27$ computed from the results in Section 4 of \cite{Riaz2006}. (Determining the onset $\lambda$ is relatively simple, however the limitations of our imaging system prohibit us from determining the onset time.)  For larger $z/\delta$, $\delta/\lambda$ decreases as the pattern coarsens and appears to asymptotically approach $\delta/\lambda\approx 0.04$. We expect this behavior for two reasons. First, if $\delta/\lambda\rightarrow 0$, then the water flux per unit interface length would tend to zero for large $z$ or $H$ which is unphysical. Second, the coarsening due to finger merging is eventually balanced by the nucleation of new fingers as the diffusion interface between two now distant fingers thickens and becomes unstable (see Fig.\ref{fig:cell}d), and for large $H$, this instability should be governed by local processes independent of $H$. After the longest finger extends to $H$, we observe some fluctuations in $\delta/\lambda$, but the time average is close to the $\delta/\lambda$ near $z/\delta=H/\delta$ in Fig.~\ref{k-A-time}a. We conclude that interactions of the fingers with the bottom of the cell do not affect the pattern, and $\delta$ is an appropriate length scale for coarsening as opposed to $H$.

Mixing of the water and PPG is important for understanding the convective process and assessing its effectiveness to securely sequester CO$_2$ in brine reservoirs. We determine the amount of water mixed from the experimental images by measuring the area $A$ between the initial interface and the middle of the subsequent diffusive interface (as in Fig.\ \ref{fig:cell}d) as a function of $t^*$\cite{Note-area}. Figure \ref{k-A-time}b shows $A/LH$ vs $t^*$ for several runs with the same $K$ but different $H$. After an initial slow increase \cite{Note-drain}, $A/LH$ increases linearly with $t^*$ indicating a steady mass transfer rate. At longer times in runs with small $H$ (e.g., the $Ra=8,700$ data), $A(t^*)/LH$ saturates due to an increase in mixture density in the lower cell which reduces $\Delta \rho$ and shuts off the convection.

\begin{figure}
\begin{center}
\includegraphics[width = 2.5 in]{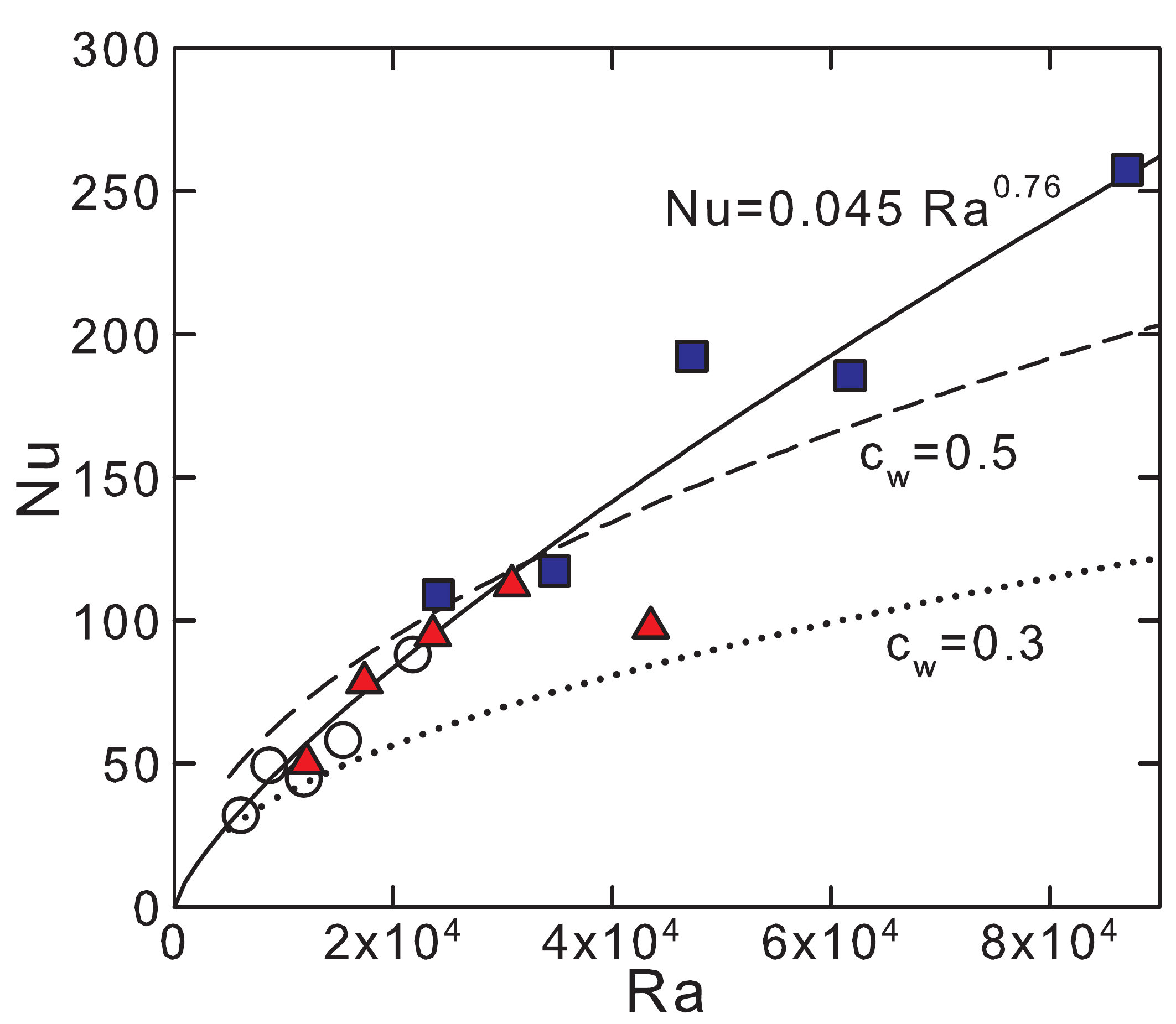}
\vspace{-0.5cm}
\end{center}
\caption{(Color online) $Nu$ vs $Ra$. Different values of $H$ are $H=1.27$ cm (open circles), $H=2.54$ cm (filled triangles), and $H=5.08$ cm (filled squares). The solid line is the best power-law fit of the form $Nu = 0.045 Ra^{0.76}$.  The dashed and dotted lines are $Nu$ computed from the simple mass flux model described in the text using the fit to $\delta/\lambda$ given in Fig 2a.  The two curves assume different values for the concentration of water in the core of the descending fingers.}
 \label{Nu-Ra}
\end{figure}

We express the mass transfer rate in dimensionless form via the Nusselt number  $Nu = \dot{m}/\rho (D/H) bL$ which represents the actual mass flux normalized by the diffusive mass flux over cell depth $H$. Rewriting, $Nu=Ra\; d(A/LH)/dt^*$, i.e. $Nu$ is simply $Ra$ times the slopes from Fig. \ref{k-A-time} which are plotted versus $Ra$ in Fig. \ref{Nu-Ra}. When fit to a power law, $Nu$ shows a dependence of $Nu=(0.045\pm 0.025) Ra^{0.76\pm0.06}$ which is similar to results from numerical simulations of thermal convection in three-dimensional porous media \cite{Otero2004} where $Nu\approx 0.017 Ra^{0.9}$ for $10^3<Ra<10^4$. Comparing values at $Ra=10,000$, we find $Nu\approx 49$ and thermal convection simulations give $Nu\approx 67$. Although the comparison with simulation is reasonable, we seek to go further by checking the internal consistency of the data in Figs.~\ref{k-A-time}a and ~\ref{Nu-Ra} and using these results to outline a new stability problem that should provide a prediction of $Nu$.

The mass flux of water through the fingers and into the lower cell is given by $\rho c_w n \delta v_f$ where $n$ is the number of fingers extending from the interface and $v_f$ and $c_w$ are the velocity and water concentration of the fluid in the finger cores. Scaling this mass flux for comparison with Fig.~\ref{Nu-Ra}, we find $Nu/Ra=c_w(\delta/\lambda)(v_f/v_c)$. To estimate $v_f/v_c$, we consider a set of equally spaced fingers and ignore small density differences except in buoyancy terms. The balance of mass flux across a horizontal line cutting through the fingers requires $v_f\delta=(\lambda-\delta)v_b$, where $v_b$ is the upward velocity of pure PPG between the fingers. Integrating the vertical component of Darcy's law around a closed rectangular contour with one vertical leg in a finger and the other in the pure PPG we find $v_f=v_c-(\nu_b/\nu_f)v_b$. Combining the previous results yields $v_f/v_c=[1+ (\nu_b/\nu_f) \delta/(\lambda-\delta)]^{-1}$ allowing us to express $Nu/Ra$ (from above) as a function solely of $\delta/\lambda$, i.e. $Nu/Ra=(\delta/\lambda)/[1+ (\nu_b/\nu_f) (\delta/\lambda)/(1-\delta/\lambda)]\equiv F[\delta/\lambda]$ ($\nu_b/\nu_f$ is fixed in this experiment). In the quasi steady state after the fingers reach $z=H$, we can instead interpret the fitted curve $\delta/\lambda=G[z/\delta]$ in Fig.~\ref{k-A-time}a as $\delta/\lambda=G[H/\delta]$. However, $H/\delta=Nu$ so that $\delta/\lambda=G[Nu]$. Combining, we find $Nu/Ra=F[G[Nu]]$ which is numerically inverted to find $Nu$ versus $Ra$ as shown in Fig.~\ref{Nu-Ra} for $c_w=0.3$ and $0.5$, i.e. $c_w$ at maximum mixture density and the maximum $c_w$ where the mixture density is still greater than $\rho_{PPG}$.

The rough agreement in Fig.~\ref{Nu-Ra} supports the validity of this simple model of mass transfer and the relationship between $Nu$ and the pattern coarsening ($\delta/\lambda$) data in Fig.~\ref{k-A-time}a. However, a full understanding of the mass transfer must include a theoretical model of $\delta/\lambda$. Here, we describe a few relevant physical process in this model, but leave the detailed derivation to future work. In our view, the core of this model includes a new stability calculation that predicts the distance between two established fingers at which the intervening diffusion layer becomes unstable to small perturbations. However, this layer is already experiencing significant transverse flows due to fluid drainage from the layer into the fingers---a significant complication not present in the instability of the original quiescent diffusion layer. Taylor dispersion due to shear along the direction of $b$ in Fig.~\ref{fig:cell}a will be quite effective in smoothing out concentration perturbations. As a result, the instability will preferentially nucleate at the symmetry point between the fingers (as is seen in Fig.~\ref{fig:cell}d and in many other occurrences of this process) where the flow stagnates and Taylor dispersion has the least effect. However, a nascent finger of finite transverse size at the stagnation point will be spread out laterally due to streamwise velocity gradients in the layer. Both of these mechanisms behave like an enhanced diffusion process which, by analogy with the quiescent diffusion layer, will increase the onset wavelength consistent with the small $\delta/\lambda$ observed in quasi steady state in Fig.~\ref{k-A-time}a.

In conclusion, our measurements of $\delta/\lambda$ at early times are consistent with models of this initial instability\cite{EnnisKing2005,Chen2006,Riaz2006} supporting the validity of these models. We have measured both the finger coarsening dynamics and steady-state convective mass transfer and outlined a new stability problem to predict the mass transfer. Our experimental results are useful for understanding the impact of reservoir parameters on the effectiveness of CO$_2$ sequestration. Permeability $K$ accounts for much of the variability in $Ra$ between different saline reservoirs. A recent survey of reservoirs in the Alberta basin\cite{Hassanzadeh2006} found a maximum $Ra\approx 1,000$. Extrapolating our best-fit power law to this lower $Ra$ yields $Nu\approx 10$. However, other reservoirs\cite{Chadwick2009} have significantly higher $K$ with $Ra\approx 20,000$ and $Nu\approx 80$. Assuming a reservoir depth of 50 m, the characteristic time for diffusive mixing is $t_d\approx 16,000$ years. At $Nu\approx 80$, the convective process observed in this work would speed up the mass transfer and reduce this time to approximately 200 years.

This work was carried out under the auspices of the
National Nuclear Security Administration of the U.S. Department of
Energy at Los Alamos National Laboratory under Contract
No. DE-AC52-06NA25396. We gratefully acknowledge the support of the U.S. Department of Energy through the LANL/LDRD Program (\#20100025DR) for this work
\bibliography{CO2}

\begin{thebibliography}{18}
\expandafter\ifx\csname natexlab\endcsname\relax\def\natexlab#1{#1}\fi
\expandafter\ifx\csname bibnamefont\endcsname\relax
  \def\bibnamefont#1{#1}\fi
\expandafter\ifx\csname bibfnamefont\endcsname\relax
  \def\bibfnamefont#1{#1}\fi
\expandafter\ifx\csname citenamefont\endcsname\relax
  \def\citenamefont#1{#1}\fi
\expandafter\ifx\csname url\endcsname\relax
  \def\url#1{\texttt{#1}}\fi
\expandafter\ifx\csname urlprefix\endcsname\relax\def\urlprefix{URL }\fi
\providecommand{\bibinfo}[2]{#2}
\providecommand{\eprint}[2][]{\url{#2}}

\bibitem[{\citenamefont{Ahlers et~al.}(2009)\citenamefont{Ahlers, Grossmann,
  and Lohse}}]{Ahlers2009review}
\bibinfo{author}{\bibfnamefont{G.}~\bibnamefont{Ahlers}},
  \bibinfo{author}{\bibfnamefont{S.}~\bibnamefont{Grossmann}},
  \bibnamefont{and} \bibinfo{author}{\bibfnamefont{D.}~\bibnamefont{Lohse}},
  \bibinfo{journal}{Rev. Mod. Phys.} \textbf{\bibinfo{volume}{81}},
  \bibinfo{pages}{503} (\bibinfo{year}{2009}).

\bibitem[{\citenamefont{Metz et~al.}(2005)\citenamefont{Metz, Davidson,
  de~Connick, Loos, and Meyer}}]{IPCC2005}
\bibinfo{author}{\bibfnamefont{D.}~\bibnamefont{Metz}},
  \bibinfo{author}{\bibfnamefont{O.}~\bibnamefont{Davidson}},
  \bibinfo{author}{\bibfnamefont{H.}~\bibnamefont{de~Connick}},
  \bibinfo{author}{\bibfnamefont{M.}~\bibnamefont{Loos}}, \bibnamefont{and}
  \bibinfo{author}{\bibfnamefont{L.}~\bibnamefont{Meyer}},
  \emph{\bibinfo{title}{IPCC special report on carbon dioxide capture and
  storage}} (\bibinfo{publisher}{Cambridge University Press},
  \bibinfo{address}{New York, NY}, \bibinfo{year}{2005}),
  chap.~\bibinfo{chapter}{5}.

\bibitem[{\citenamefont{{Foster}}(1965)}]{Foster65}
\bibinfo{author}{\bibfnamefont{T.~D.} \bibnamefont{{Foster}}},
  \bibinfo{journal}{Physics of Fluids} \textbf{\bibinfo{volume}{8}},
  \bibinfo{pages}{1249} (\bibinfo{year}{1965}).

\bibitem[{\citenamefont{{Ennis-King} et~al.}(2005)\citenamefont{{Ennis-King},
  {Preston}, and {Paterson}}}]{EnnisKing2005}
\bibinfo{author}{\bibfnamefont{J.}~\bibnamefont{{Ennis-King}}},
  \bibinfo{author}{\bibfnamefont{I.}~\bibnamefont{{Preston}}},
  \bibnamefont{and}
  \bibinfo{author}{\bibfnamefont{L.}~\bibnamefont{{Paterson}}},
  \bibinfo{journal}{Physics of Fluids} \textbf{\bibinfo{volume}{17}},
  \bibinfo{pages}{084107} (\bibinfo{year}{2005}).

\bibitem[{\citenamefont{{Xu} et~al.}(2006)\citenamefont{{Xu}, {Chen}, and
  {Zhang}}}]{Chen2006}
\bibinfo{author}{\bibfnamefont{X.}~\bibnamefont{{Xu}}},
  \bibinfo{author}{\bibfnamefont{S.}~\bibnamefont{{Chen}}}, \bibnamefont{and}
  \bibinfo{author}{\bibfnamefont{D.}~\bibnamefont{{Zhang}}},
  \bibinfo{journal}{Advances in Water Resources} \textbf{\bibinfo{volume}{29}},
  \bibinfo{pages}{397} (\bibinfo{year}{2006}).

\bibitem[{\citenamefont{{Riaz} et~al.}(2006)\citenamefont{{Riaz}, {Hesse},
  {Tchelepi}, and {Orr}}}]{Riaz2006}
\bibinfo{author}{\bibfnamefont{A.}~\bibnamefont{{Riaz}}},
  \bibinfo{author}{\bibfnamefont{M.}~\bibnamefont{{Hesse}}},
  \bibinfo{author}{\bibfnamefont{H.~A.} \bibnamefont{{Tchelepi}}},
  \bibnamefont{and} \bibinfo{author}{\bibfnamefont{F.~M.} \bibnamefont{{Orr}}},
  \bibinfo{journal}{Journal of Fluid Mechanics} \textbf{\bibinfo{volume}{548}},
  \bibinfo{pages}{87} (\bibinfo{year}{2006}).

\bibitem[{\citenamefont{Ennis-King and Paterson}(2007)}]{EnnisKing2007}
\bibinfo{author}{\bibfnamefont{J.}~\bibnamefont{Ennis-King}} \bibnamefont{and}
  \bibinfo{author}{\bibfnamefont{L.}~\bibnamefont{Paterson}},
  \bibinfo{journal}{International Journal of Greenhouse Gas Control}
  \textbf{\bibinfo{volume}{1}}, \bibinfo{pages}{86 } (\bibinfo{year}{2007}),
  ISSN \bibinfo{issn}{1750-5836}.

\bibitem[{\citenamefont{{Rapaka} et~al.}(2008)\citenamefont{{Rapaka}, {Chen},
  {Pawar}, {Stauffer}, and {Zhang}}}]{Rapaka2008}
\bibinfo{author}{\bibfnamefont{S.}~\bibnamefont{{Rapaka}}},
  \bibinfo{author}{\bibfnamefont{S.}~\bibnamefont{{Chen}}},
  \bibinfo{author}{\bibfnamefont{R.~J.} \bibnamefont{{Pawar}}},
  \bibinfo{author}{\bibfnamefont{P.~H.} \bibnamefont{{Stauffer}}},
  \bibnamefont{and} \bibinfo{author}{\bibfnamefont{D.}~\bibnamefont{{Zhang}}},
  \bibinfo{journal}{Journal of Fluid Mechanics} \textbf{\bibinfo{volume}{609}},
  \bibinfo{pages}{285} (\bibinfo{year}{2008}).

\bibitem[{\citenamefont{{Rapaka} et~al.}(2009)\citenamefont{{Rapaka}, {Pawar},
  {Stauffer}, {Zhang}, and {Chen}}}]{Rapka2009}
\bibinfo{author}{\bibfnamefont{S.}~\bibnamefont{{Rapaka}}},
  \bibinfo{author}{\bibfnamefont{R.~J.} \bibnamefont{{Pawar}}},
  \bibinfo{author}{\bibfnamefont{P.~H.} \bibnamefont{{Stauffer}}},
  \bibinfo{author}{\bibfnamefont{D.}~\bibnamefont{{Zhang}}}, \bibnamefont{and}
  \bibinfo{author}{\bibfnamefont{S.}~\bibnamefont{{Chen}}},
  \bibinfo{journal}{Journal of Fluid Mechanics} \textbf{\bibinfo{volume}{641}},
  \bibinfo{pages}{227} (\bibinfo{year}{2009}).

\bibitem[{\citenamefont{Kneafsey and Pruess}(2010)}]{Pruess2010}
\bibinfo{author}{\bibfnamefont{T.}~\bibnamefont{Kneafsey}} \bibnamefont{and}
  \bibinfo{author}{\bibfnamefont{K.}~\bibnamefont{Pruess}},
  \bibinfo{journal}{Transport in Porous Media} \textbf{\bibinfo{volume}{82}},
  \bibinfo{pages}{123} (\bibinfo{year}{2010}), ISSN \bibinfo{issn}{0169-3913},
  \bibinfo{note}{10.1007/s11242-009-9482-2}.

\bibitem[{dow()}]{dowppgdata}
\bibinfo{note}{Density and dynamic viscosity data for propylene glycol are
  taken from manufacturer data, http://dow-answer.custhelp.com/}.

\bibitem[{\citenamefont{Hubel et~al.}(2002)\citenamefont{Hubel, Bidault, and
  Hammer}}]{D_ppg}
\bibinfo{author}{\bibfnamefont{A.}~\bibnamefont{Hubel}},
  \bibinfo{author}{\bibfnamefont{N.}~\bibnamefont{Bidault}}, \bibnamefont{and}
  \bibinfo{author}{\bibfnamefont{B.}~\bibnamefont{Hammer}},
  \bibinfo{journal}{ASME Conference Proceedings}
  \textbf{\bibinfo{volume}{2002}}, \bibinfo{pages}{121} (\bibinfo{year}{2002}).

\bibitem[{Not({\natexlab{a}})}]{Note-drain}
\bibinfo{note}{The shim addenda volume produces a delay in the motion of the
  interface estimated to be of order of several $t^*$.}

\bibitem[{\citenamefont{{Taylor}}(1953)}]{Taylor1953}
\bibinfo{author}{\bibfnamefont{G.}~\bibnamefont{{Taylor}}},
  \bibinfo{journal}{Royal Society of London Proceedings Series A}
  \textbf{\bibinfo{volume}{219}}, \bibinfo{pages}{186} (\bibinfo{year}{1953}).

\bibitem[{Not({\natexlab{b}})}]{Note-area}
\bibinfo{note}{There are corrections of order 5\% due to the small density
  differences between the fluids.}

\bibitem[{\citenamefont{{Otero} et~al.}(2004)\citenamefont{{Otero},
  {Dontcheva}, {Johnston}, {Worthing}, {Kurganov}, {Petrova}, and
  {Doering}}}]{Otero2004}
\bibinfo{author}{\bibfnamefont{J.}~\bibnamefont{{Otero}}},
  \bibinfo{author}{\bibfnamefont{L.~A.} \bibnamefont{{Dontcheva}}},
  \bibinfo{author}{\bibfnamefont{H.}~\bibnamefont{{Johnston}}},
  \bibinfo{author}{\bibfnamefont{R.~A.} \bibnamefont{{Worthing}}},
  \bibinfo{author}{\bibfnamefont{A.}~\bibnamefont{{Kurganov}}},
  \bibinfo{author}{\bibfnamefont{G.}~\bibnamefont{{Petrova}}},
  \bibnamefont{and} \bibinfo{author}{\bibfnamefont{C.~R.}
  \bibnamefont{{Doering}}}, \bibinfo{journal}{Journal of Fluid Mechanics}
  \textbf{\bibinfo{volume}{500}}, \bibinfo{pages}{263} (\bibinfo{year}{2004}).

\bibitem[{\citenamefont{Hassanzadeh et~al.}(2006)\citenamefont{Hassanzadeh,
  Pooladi-Darvish, and Keith}}]{Hassanzadeh2006}
\bibinfo{author}{\bibfnamefont{H.}~\bibnamefont{Hassanzadeh}},
  \bibinfo{author}{\bibfnamefont{M.}~\bibnamefont{Pooladi-Darvish}},
  \bibnamefont{and} \bibinfo{author}{\bibfnamefont{D.}~\bibnamefont{Keith}},
  \bibinfo{journal}{Transport in Porous Media} \textbf{\bibinfo{volume}{65}},
  \bibinfo{pages}{193} (\bibinfo{year}{2006}), ISSN \bibinfo{issn}{0169-3913},
  \bibinfo{note}{10.1007/s11242-005-6088-1}.

\bibitem[{\citenamefont{Chadwick et~al.}(2009)\citenamefont{Chadwick, Noy,
  Arts, and Eiken}}]{Chadwick2009}
\bibinfo{author}{\bibfnamefont{R.}~\bibnamefont{Chadwick}},
  \bibinfo{author}{\bibfnamefont{D.}~\bibnamefont{Noy}},
  \bibinfo{author}{\bibfnamefont{R.}~\bibnamefont{Arts}}, \bibnamefont{and}
  \bibinfo{author}{\bibfnamefont{O.}~\bibnamefont{Eiken}},
  \bibinfo{journal}{Energy Procedia} \textbf{\bibinfo{volume}{1}},
  \bibinfo{pages}{2103 } (\bibinfo{year}{2009}), ISSN
  \bibinfo{issn}{1876-6102}.

\end{thebibliography}

\end{document}